\documentclass[preprint, 5p, times,twocolumn]{elsarticle}

\usepackage{flushend}
\usepackage{threeparttable}
\usepackage{hyperref,latexsym}
\usepackage{graphicx}
\usepackage{amsmath,siunitx}
\usepackage{amssymb}
\usepackage{soul}
\usepackage{color, xcolor}
\usepackage{booktabs}
\usepackage{upgreek}
\upshape
\usepackage{overpic}
\usepackage{physics,bm,xfrac,nicefrac}
\usepackage{mathtools}
\usepackage{todonotes} % provides commands \delete etc.
 \usepackage{float}

\DeclareMathOperator{\E}{\mathbb{E}}

\DeclareMathOperator{\Kur}{\mathcal{K}}
\DeclareMathOperator{\Pdf}{\mathbb{P}}

\graphicspath{{fig/}{figures/}}

\begin{document}

% \input{aip-titlepage}
%\bibliographystyle{elsarticle-num}
%\journal{Nucl. Instrum. Meth. A}
\journal{}

\begin{frontmatter}
    % \title{Mometum-revised  muon scattering tomography \\ The momentum effects in muon tomography image reconstruction }
    \title{ Improving Muon Scattering Tomography Performance With A Muon Momentum Measurement Scheme }

    \author[a,b]{Pei Yu\corref{cor1}}
    \cortext[cor1]{These authors contributed equally to this work.}  
    \author[c,d]{Ziwen Pan\corref{cor1}}
    \author[a,b]{Jiajia Zhai}
    \author[a,b]{Yu Xu}
    \author[a,b]{Li Deng}
    \author[c,d]{Zhengyang He}
    \author[c,d]{Zhe Chen}
    \author[c,d]{Zechao Kang}
    \author[a,b,e,f]{Yuhong Yu}
    \author[a,b,e,f]{Xueheng Zhang}
    \author[a,b,e,f]{Liangwen Chen\corref{cor2}}
    \cortext[cor2]{Corresponding author}
    \ead{chenlw@impcas.ac.cn}
    
    \author[a,b,e,f]{Lei Yang}        
    \author[a,b,e,f]{Zhiyu Sun}

    \address[a]{Advanced Energy Science and Technology Guangdong Laboratory, Huizhou 516227, China}
    \address[b]{Institute of Modern Physics, Chinese Academy of Sciences, Lanzhou 730000, China}
    \address[c]{State Key Laboratory of Particle Detection and Electronics, University of Science and Technology of China, Hefei 230026, China}
    \address[d]{Department of Modern Physics, University of Science and Technology of China, Hefei 230026, China}
    \address[e]{School of Nuclear Science and Technology, University of Chinese Academy of Sciences, Beijing 100049, China}
    \address[f]{State Key Laboratory of Heavy Ion Science and Technology, Institute of Modern Physics, Chinese Academy of Sciences, Lanzhou 730000, China}
 
\begin{abstract}
Muon imaging, especially muon scattering tomography (MST),  has recently garnered significant attention.
MST measures the magnitude of muon scattering angles inside an object, which depends not only on the material properties but also on the muon momentum.
Due to the difficulty of simultaneous measurement of momentum, it was neglected and taken as a constant in multiple MST reconstruction algorithms.
Recently, an experimental measurement scheme has emerged that is feasible in engineering, but it requires many layers of detectors to approach the true momentum.
From this, we proposed both an algorithm to incorporating momentum into MST, and a scheme to determine the thresholds of Cherenkov detectors.
This novel scheme, termed the "equi-percentage scheme", sets momentum thresholds for Cherenkov detector layers based on cosmic muon momentum distribution.
Results showed our approach delivers noticeable enhancement in reconstructed image quality even with only two detector layers, reaching near-saturation performance with four layers. This study proves that momentum measurement significantly enhances short-duration MST, and that substantial improvement can be achieved with relatively coarse momentum measurement using 2-4 layers of Cherenkov detectors.
\end{abstract}
\end{frontmatter}

%  \linenumbers

\section{Introduction}
\label{sec:intro}

Muon tomography is a non-invasive technique that utilizes muons to image the internal structure of
large-scale objects, such as volcanoes \cite{tanaka2019,Tioukov2019}, pyramids
\cite{Alvarez:1970ecc,Morishima:2017ghw}, and nuclear safety and security \cite{Errico:2020lgx}, etc. These muon tomography applications rely on the exceptional penetrating capability of muons.
Unlike electrons and protons, muons do not undergo strong interactions, and thus exhibit much less energy loss when passing through matter, rendering them great penetrating ability. At the Earth surface, the cosmic-ray muon are characterized by an average energy of several GeVs and a flux of
approximately 1 $\text{cm}^{-1}\text{min} ^{-1}$~\cite{Durham:2017zfw}. Additionally, our recent work \cite{Xu:2025spd} showed that GeV energy muons could be available in High Intensity Heavy-Ion Accelerator Facility (HIAF),  with an intensity reaching $\sim 10^{6}$ muons/s. These high-energy muons penetrate dense materials with scattering or absorption that depends on the material composition and density. Therefore, the density map of materials can be obtained by measuring muon flux attenuation or muon scattering.

In muon scattering tomography (MST), the Coulomb scattering information between the muons and the
materials is used to image the objects. When passing through materials, the angular distribution of
scattered muons can be well approximated by a Gaussian distribution, with its characteristic width
directly correlated with both the material properties and the incident muon momentum
\cite{Schultz:2004kx,Yu:2024mt}.  Since natural cosmic-ray muons exhibit a continuous energy
spectrum extending up to TeV-scale momenta, real-time momentum measurement proves extremely
challenging.  Therefore, most studies have typically assumed a fixed momentum value of $3 - 4$ GeV/c for all incident muons, corresponding to the average cosmic-ray muon momentum at sea level
\cite{Xiao:2018,Bae2024}. However, this simplified approximation may prove inadequate for
applications demanding high spatial resolution\cite{Durham2018,Li2023}. 

In order to address this limitation, researchers have developed several mitigation approaches. As
demonstrated in Ref. \cite{Xiao:2018}, implementing a modified multi-group model that accounts for
both the angular and momentum distributions of cosmic-ray muons can effectively improve image
resolution of muon tomography in practical applications. The study in Ref. \cite{osti_23047380}
revealed that the incorporation of muon momentum shows particular promise for improving the
reconstruction resolution of cask imaging.  In Ref. \cite{Luo2015}, researchers exploited the
potential of using multi-gap resistive plate chambers (MRPCs) to obtain segmented muon energy
information using time-of-flight (ToF) measurements. Their experimental results have shown that the ToF energy segmentation performs a $25\%$ enhancement in the identification of high Z materials than those
without muon energy.  Ref. \cite{Vanini2018} proposed a method to estimate the muon momentum by
means of the scattering in the detector's material.  Ref. \cite{Li2023} adopted several discrete
muon energies to fit the scattering angle distribution of cosmic ray muons and estimate the weight of each discrete muon energy. Their analysis shows that the incorporation of discrete muon energy
information in this way can effectively improve the identification accuracy of high Z nuclear
materials.  Ref. \cite{Chen2023} developed a new MST system that integrates Cherenkov detectors,
which utilize momentum measurement of low-energy muons, to discriminate both low-Z and high-Z
materials. They found that measuring the momentum of low energy muons can significantly improve the
identification power of low-Z materials.  In Refs. \cite{Bae2022,Bae2024,Bae2024I}, investigators had explored a concept of measuring muon momentum by using a multi-layer gas Cherenkov spectrometer as well as a momentum dependent imaging algorithm for imaging reconstruction. The significant
benefits of muon momentum incorporation in improving the resolution of muon tomography was demonstrated.

Since the muon momentum measurement has the potential in enhancing the muon tomography resolution,
this paper proposes an algorithm that integrates momentum information into the image reconstruction processes and explores the achievable imaging performance with the new algorithm under limited momentum resolution. Furthermore, a scheme for a multilayer Cherenkov detector spectrometer is raised in this work, with a great promotion in image quality even using only two detector layers, making it more feasible to experimentally measure the momentum in MST.

This paper is organized as follows. Section~\ref{sec:methods} introduces the tomography method to
incorporate the momentum information, as well as the elimination of bright spots. Section~\ref{sec:results:1} shows the imaging results, and discusses the improvement of imaging quality after
incorporating the muon momentum and removing bright spots in reconstructed images. In Sections~\ref{sec:result:MIS} and \ref{sec:result:MIS_custom}, two momentum discretion methods are proposed for
the multi-layer Cherenkov detector spectrometer, respectively. Their influences on imaging
reconstruction quality are studied and compared. Section~\ref{sec:discussion} quantitatively
discusses the aid of muon momentum in improving the MST imaging quality. Section~\ref{sec:conclude} summaries the simulation study of the new momentum incorporation in MST imaging reconstruction.

\section{Methods}\label{sec:methods}

When muons traverse a matter, multiple Coulomb scattering occurs.
Given the muon momentum, $p$, the scattering of a muon in the three-dimensional space can be decomposed into two mutually perpendicular planes which contain the incident line, with the scattering angle in each plane approximately following a Gaussian distribution~\cite{Bethe:1953va}
\begin{align}
    \Pdf(\theta) = \frac{1}{\sqrt{2\pi}\sigma} \exp\left(-\frac{\theta^2}{2 \qty(\frac{15\,\text{MeV}}{\beta cp})^2\frac{\Delta L}{L_\text{rad}}}\right)\,,
    % P(\theta) &= \frac{1}{\sqrt{2\pi}\sigma} \exp\left(-\frac{\theta^2}{2\sigma^2\frac{\Delta L}{L_\text{rad}}}\right)\,, \\
    % \sigma    &= \frac{15\,\text{MeV}}{\beta cp} \,,
    \label{eq:muon_gaussian}
\end{align}
where $\Pdf$ stands for the probability, $\theta$ is the scattering angle in each plane, $\beta c$ is
the muon's relativistic velocity, $p$ is the muon momentum, $\sigma$ is the standard deviation, $\Delta L$ is the thickness of the object, and the $L_\text{rad}$ is the radiation length in the object. The $L_\text{rad}$ depends on the type and density of the material in the object, which can be expressed as
\begin{align}
    L_{\text{rad}} &= \frac{(\SI{716.4}{g\cdot.cm^{-2})\cdot A}}{\rho{}Z(Z+1)\ln(287/\sqrt{Z})}\,,
\end{align}
% \todo[inline]{公式的解释要写}
% 式中 $\beta c$ 与 $p$ 分别为缪子的相对论速度和动量，$\Delta L$是散射物体的厚度，$L_\text{rad}$ 为辐射长度。辐射长度与物体的性质相关，如下式\cite{Morishima:2017ghw}
where $A$ and $Z$ represent the mass number and atomic number of the material, respectively, while $\rho$ denotes the material density.
For three-dimensional imaging, we aim to define an imaging density that is independent of the voxel size division. This can be expressed as:
\begin{align}
    \lambda \equiv \frac{\sigma^2}{\Delta L} 
    \approx \qty(\frac{\SI{15}{MeV}}{c\beta p})^2\frac{\rho{}Z(Z+1)\ln(287/\sqrt{Z})}{\qty(\SI{716.4}{g.cm^{-2} })\cdot A}\,,
    \label{eq:lambda_def}
\end{align}
\begin{align}
    \sigma^2 &\approx \qty(\frac{15\,\text{MeV}}{\beta cp})^2\frac{\Delta L}{L_\text{rad}},
    \label{eq:muon_sigma}
\end{align}
where $\sigma$ is the variance in the angular distribution of muon scattering in Eq.~\eqref{eq:muon_gaussian}.
It shows that the influence of the object thickness is eliminated, and its magnitude depends only on
the material type and the muon momentum. In a practical imaging, the distribution width of the detected scattering angle is used, under certain approximations, to inversely determine $\lambda$ in
Eq.~\eqref{eq:lambda_def}.

\subsection{Imaging methods}
\label{sec:methods:poca}

The Point of Closest Approach (PoCA) method~\cite{Schultz:2004kx} is an algorithm proposed by Los Alamos National Laboratory when introducing the concept of muon scattering imaging. It serves as the foundation for later scattering imaging algorithms.
In MST experiments, the scattering angle is determined by the detection of the incoming and outgoing trajectories of the muon. This requires two groups of detectors with the object placed between them. Each detector group comprises at least two position-sensitive detectors with a moderate gap to measure muon tracks that are fitted as lines.
% After the incoming line and outgoing line are determined,
The PoCA point\cite{Schultz:2004kx} is defined as the midpoint of the common perpendicular line between the fitted incoming and outgoing tracks. The actual scattering path of muons in an object is curved and cannot be directly measured. In the PoCA algorithm, the multiple scattering process is approximated as a single scattering event occurring at the PoCA point. It means that the incoming and outgoing muon tracks measured by two detector groups are utilized to approximate muon trajectories inside the sample.

% In simpler imaging scenarios, the spatial distribution of PoCA points is used to generate a three-dimensional image.
The imaging space is divided into voxels for image reconstruction. In an experiment with many muon counts, every voxel is populated with several events. The imaging density in every voxel is defined as the standard deviation of the scattering angle distribution contributed by the recorded events. 
For these events, the fitted muon tracks pass through the voxel. However, it is possible that the
reconstructed PoCA point is outside the voxel. If so, the scattering angle in the image density
calculation is set to 0 ($\theta_{ij} = 0$);
otherwise, it is the scattering angle of this event ($\theta_{ij} = \theta_i$).
Accordingly, the imaging density is determined by
\begin{align}
    \lambda_j = \frac{\sigma^2}{\Delta L} =  \frac{\frac{1}{\abs{E_j}}\sum_{i \in E_j }\theta_{ij}^2}{\Delta L_\text{voxel}}\,.
    \label{eq:lambda_2}
\end{align}
In this equation, the subscript $j$ denotes the $j$-th voxel, $E_j$ represents
all events passing through the $j$-th voxel with the total number of events denoted as $\abs{E_j}$, $i$ is the index of muon scattering events recorded in $E_j$, and $\theta_{ij}$ represents the scattering angle of $i$-th event within the $j$-th voxel.

Since the number of events in a voxel is proportional to the total cumulative path length of muons within it, we reformulate Eq.~(\ref{eq:lambda_2}) as
\begin{align}
    \lambda_j = \frac{\Theta_j}{L_j}, \quad
    \Theta_j \equiv \sum_{i \in E_j   }  \theta_{ij}^2, \quad
    L_j \equiv \sum_{i \in E_j  }  \Delta L_{ij}\,.
    \label{eq:lambda_3}
\end{align}
Herein, $\Theta_j$ represents the sum of squared scattering angles from all
muons passing through the $j$-th voxel, while $L_j$ denotes the total path
length traversed by all muons within this voxel. This formulation is
computationally advantageous for our implementation and serves as the
foundation for a subsequent big-voxel processing algorithm which was proposed in the previous work\cite{Yu:2024mt}.

The big-voxel algorithm first partitions the
space into smaller voxels, accumulating the aforementioned $\Theta_j$ and $L_j$
information within each voxel. When calculating the imaging density value, the
algorithm combines $\Theta_j$ and $L_j$ from all surrounding voxels that can be taken as a big voxel. The same operation is employed for every small voxel. Therefore, the statistical fluctuation of scattering angles in a small voxel is greatly suppressed using the big-voxel processing.

\subsection{Momentum corrected tomography}
\label{sec:methods:mome}

% In previous muon scattering imaging, we were unable to measure the momentum of each scattering event. 
Equations (\ref{eq:muon_gaussian}-\ref{eq:muon_sigma}) work for muons with a given energy.
However, in cosmic-ray muon imaging experiment, the muon energy distributes widely from sub-GeVs to
TeVs. Due to the difficulties in the simultaneous measurements of muon momentum, 
Eqs.~(\ref{eq:lambda_def}-\ref{eq:muon_sigma}) are still utilized with the cosmic-ray muon momentum fixed to a
certain value. Such a treatment can still reconstruct the density distribution of materials in an
object.
% Consequently, the reconstructed image exhibits a certain degree of blurring.
As shown in Eq.~\eqref{eq:muon_sigma}, the scattering angle distribution width is inversely proportional to the muon momentum.
The mismatch of assigned muon momentum and measured scattering angles leads to worse image quality.
% \pdfcomment{模糊不是因为这个}
This will be more evident for those detected muon events with lower momenta. If the muon momentum could be  properly measured and incorporated into the image reconstruction process, the imaging quality could be improved due to the availability of additional information.

We incorporate momentum information as follows. The scattering angle $\theta$ is corrected to $\theta'$ by employing
\begin{align}
    \theta &\rightarrow \theta' \equiv \theta\, \frac{p}{p_0},
    \label{eq:tau_def}
\end{align}
where $p_0$ is a constant muon momentum (typically the average
momentum) to keep the scattering angle in the same unit system after corrections. The addition of $p_0$ has no influence on the relative brightness of a reconstructed image. Thus, the corrected imaging density, $\lambda'$, can be expressed as
\begin{align}
    % P(\tau) &= \frac{1}{\sqrt{2\pi}\sigma} \exp\left(-\frac{\tau^2}{2{(\sigma p)}^2}\right) \\
    % \lambda'_j& = \frac{\sigma^2}{\Delta L} =  \frac{\frac{1}{\abs{E_j}}\sum_{i=1}^{\left | E_j \right | }  \tau_{ij}^2}{\Delta L_\text{voxel}},
    \lambda'_j& = \frac{\sigma^2}{\Delta L} =  \frac{\frac{1}{\abs{E_j}}\sum_{i\in E_j}  \theta_{ij}
    ^{'2}}{\Delta L_\text{voxel}} = \frac{\frac{1}{\abs{E_j}}\sum_{i\in E_j}  (\theta_{ij}\frac{p_i}{p_0})^{2}}{\Delta L_\text{voxel}} \,.
    \label{eq:lambda_prime}
\end{align}
Herein, $p_i$ represents the muon momentum of $i$-th event within the $j$-th voxel. This new density formula will have better performance than the origin density. Moreover, in MST experiments, the muon momentum measuring precision is limited. This paper will discuss the momentum measuring scenario selections and their influences on imaging quality. 

\subsection{Simulation setup}

\begin{figure}
	\centering
	\includegraphics[width=0.5\textwidth]{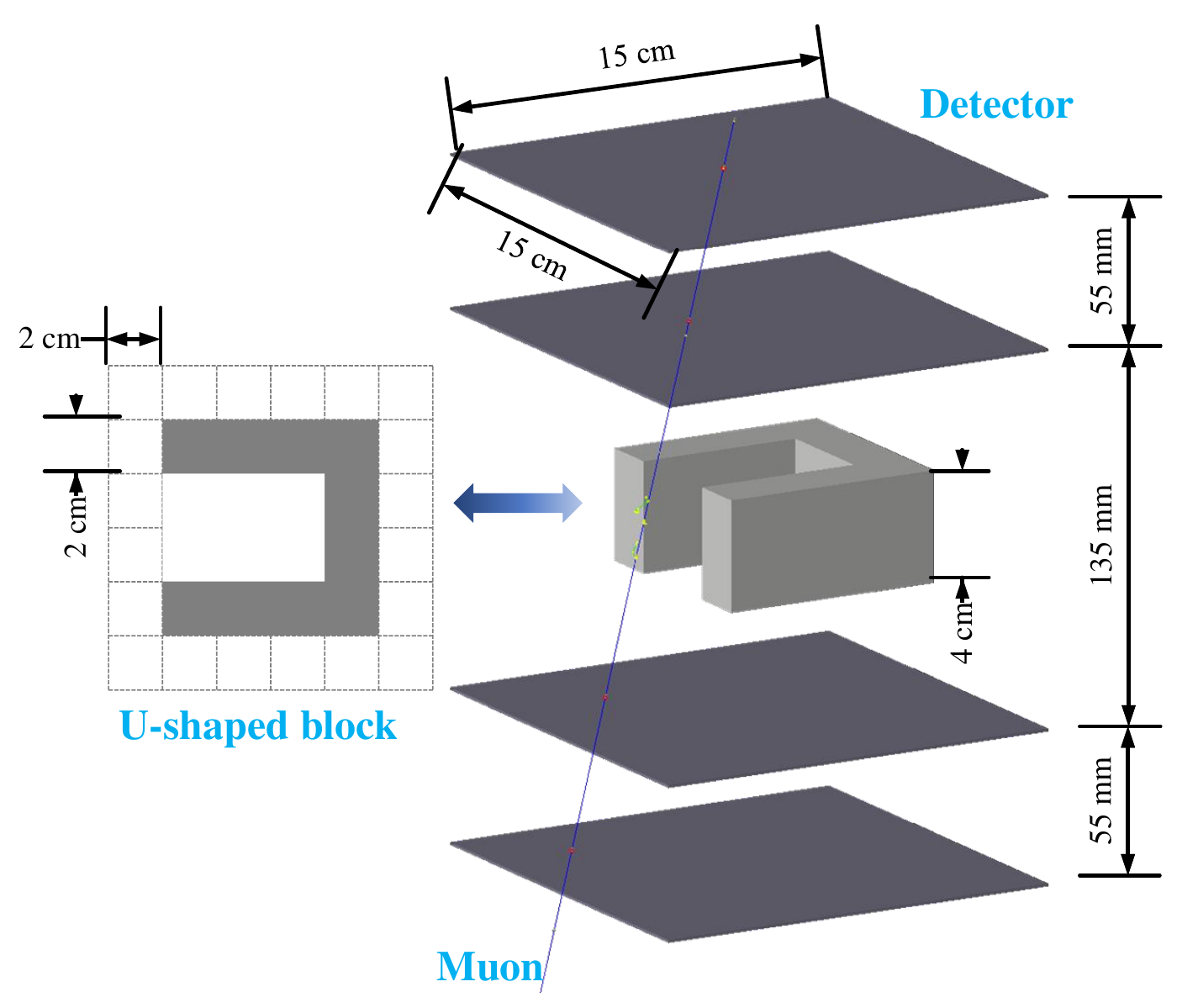}
	\caption{Illustration of the simulation model. The setup consists of two detector groups with each one composed of a  pair of square detectors in dimensions of \qtyproduct{15x15}{cm}. The imaging object is an U-shape tungsten block.
		%\deleted{Sizes are not to scale for clear display.} 这个图就是Geant4截图，是真实比例
	} \label{fig:sim-setup}
\end{figure}

The MST data in this study are obtained from Monte Carlo simulations. Herein, the Geant4 toolkit
\cite{agostinelli2003geant4,allison2006geant4,guatelli2011introduction} is employed to simulate the
physical processes of muons penetrating a U-shaped object composed of tungsten as shown in Fig.~\ref{fig:sim-setup}. Two groups of virtual detectors are placed above and below the block to
record the incoming and outgoing tracks of muons, respectively. Sizes are indicated in the
illustrative diagram. The  Cosmic-ray Shower Library (CRY) \cite{CRY:Hagmann2007} is adopted to
sample the energy and emission direction of those muons. The muon momentum is recorded when
penetrating the top detector, and it is used in the momentum correction. The `random seed' is uniquely set for every simulation. In short-term imaging, the image quality may be affected by the random number seed. Therefore, the imaging quality in this paper is calculated in multiple simulations with different dedicated random number seeds, to eliminate the influence of the seed selection on the results.

\subsection{Image quality assessment}

This study employs the Structural Similarity Index Measure (SSIM) index\cite{Wang2004:SSIM} as the quantitative metric to evaluate the quality of reconstructed images. The SSIM index measures the similarity between a test image and a reference image, with values approaching 1 indicating higher similarity. The mathematical formulation is expressed as
\begin{align}
    \operatorname{SSIM}(a,b) = \frac{ (2\mu_a \mu_b + C_1) }{ (\mu_a^2 + \mu_b^2 + C_1) }
    \cdot
    \frac{ (2\sigma_{\!ab} + C_2) }{ (\sigma_a^2 + \sigma_b^2 + C_2) }\,,
    \label{eq:SSIM}
\end{align}
where $(a,b)$ denote the two images being compared, $\mu$ and $\sigma^2$ represent the mean and variance of a pixel, respectively, and $\sigma_{\!ab}$ is the covariance between corresponding pixels in both images. The constants, $C_1=0.01$ and $C_2=0.03$, serve to stabilize the denominator while maintaining consistency with the original SSIM implementation \cite{Wang2004:SSIM}.
 
\section{Results}\label{sec:results}

\subsection{Momentum-corrected imaging, bright-spot removal, and their SSIM comparisons} \label{sec:results:1}

\begin{figure}
	    \centering
	\includegraphics[width=0.49\textwidth]{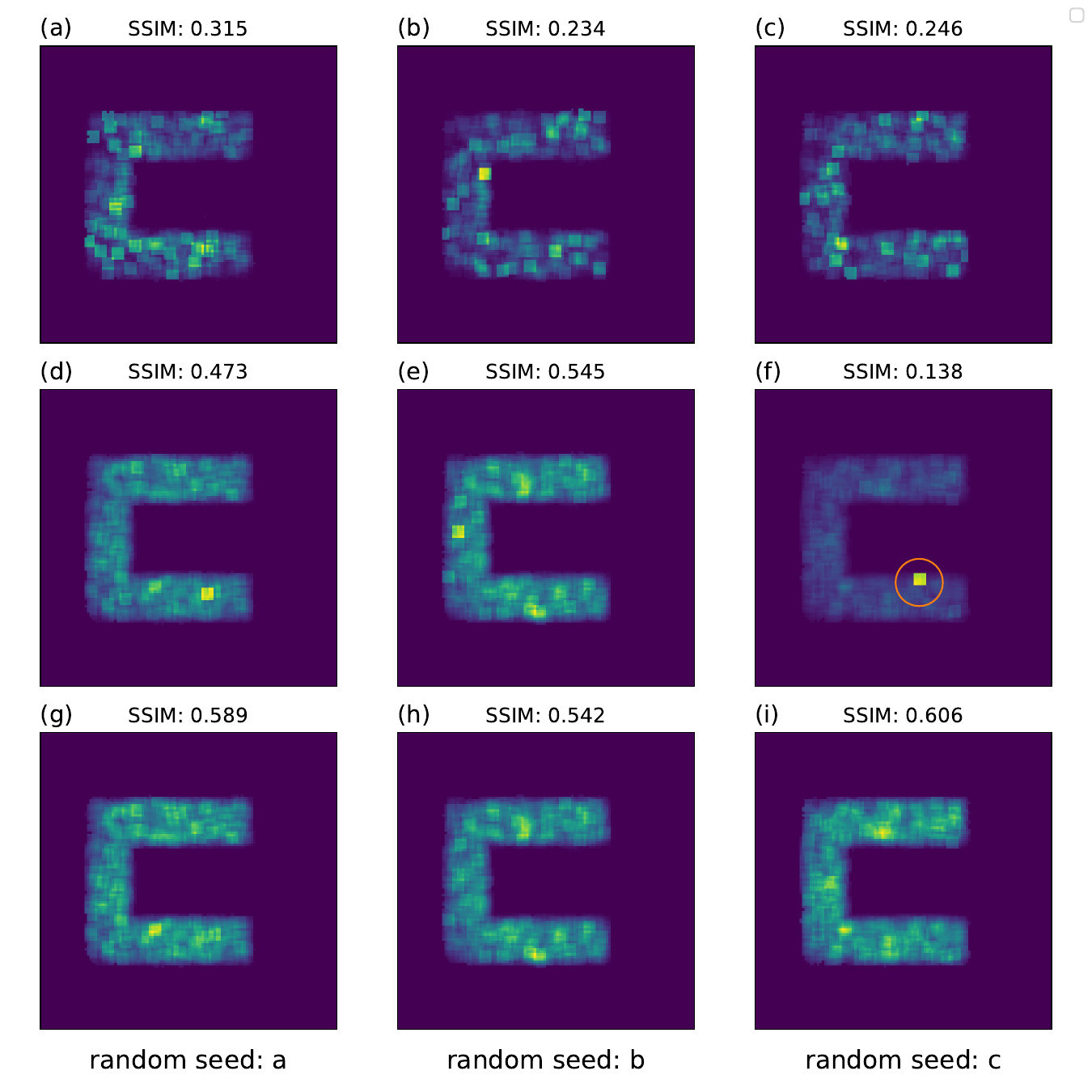}
	\caption{ Comparison of reconstructed images: (a‑c) no momentum correction and no bright spots removal, (d‑f) with momentum correction and no bright spots removal, (g‑i) with momentum correction and bright spots removal. Each column corresponds to the same simulation run identified by the same random seed. A number of $2\times10^4$ muons were recorded by four detectors.}
	\label{fig:result1}
\end{figure}

Fig.~\ref{fig:result1} presents reconstructed images with or without momentum corrections and bright spots removals. The influence of random number sampling is compared by columns. A number of $2\times10^{4}$ events were simulated in each run, equivalent to approximately 24 hours of exposure time. Comparing Fig.~\ref{fig:result1}(a‑c) and (d-e), the SSIM indices of reconstructed images become more uniform and much better after momentum corrections. However, a localized bright spot randomly occurs in Fig.~\ref{fig:result1}(f), causing the rest of the image to appear relatively dim and resulting in a lower SSIM score (0.138). Although the structural details of the object remain clearly visible in the less intense regions, the presence of a bright spot significantly degrades the SSIM index. Thus, removing these outliers is essential to accurately evaluate the true structural representation of a reconstructed image.

We employ the following processes to remove bright spots. As described in 
Eqs.~(\ref{eq:lambda_3}-\ref{eq:tau_def}), the squared scattering angles (or corrected scattering
angles) in a voxel are accumulated in the variable $\Theta$. The distribution of corrected $\theta'$ can be histogrammed as shown in Fig.~\ref{fig:angle-distribution}(b). The circled data point in this graph is corresponded to the bright spot marked in Fig.~\ref{fig:result1}(f). To suppress the influence of the bright spot, the $\theta'$ values over the maximum cut are set the same. Herein, the maximum cut is defined at the $N$-th maximum $\theta'$ value. In this work, $N$ is set to 10 as less than 4 bright spots existing almost in all reconstructed images. The image will be reconstructed afterwards. Considering that an reconstructed image contains at least ($100\times 100 = 10000$) pixels, the bright spot removing treatment affects only  $N$ pixels. Thus, this treatment has negligible impacts on the whole image. Fig.~\ref{fig:result1}(g-i) employ both the momentum correction and bright spot removal treatments. The improvement of image definition and SSIM indices demonstrated the effectiveness of the bright spot removal processing.

\begin{figure}
	\centering
	\includegraphics[width=0.5\textwidth]{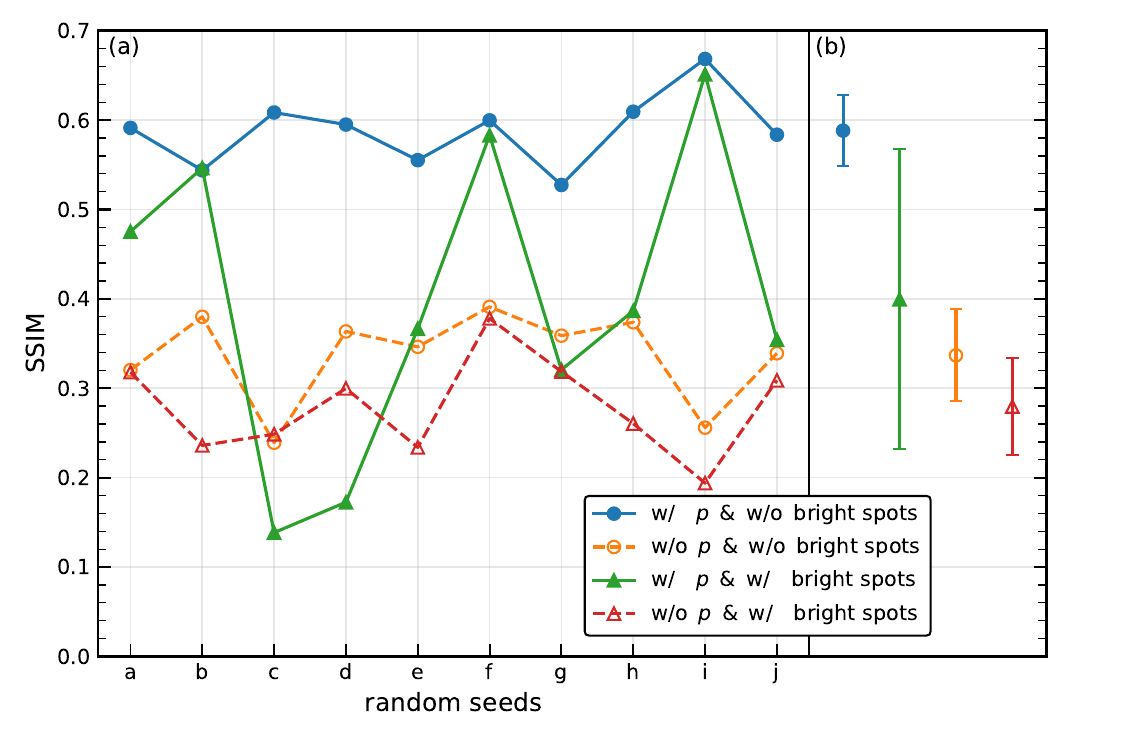}
	\caption{ (a) The SSIM indices for combinations of whether or not the momentum correction or the bright spot removal is applied. Labels ``w/'' denotes ``with'', while ``w/o'' denotes ``without''. A number of $2\times10^4$ muons were recorded that every muon penetrated four detectors simultaneously. For every dataset, the simulation was repeated 10 times with the random seeds marked from a to j. (b) The mean and error bar for every dataset shown in panel (a).}
	\label{fig:ssim-with-mome}
\end{figure}

Beyond those three representative cases in Fig.~\ref{fig:result1}, we systematically evaluated the SSIM indices across multiple
random seeds, with and without momentum corrections as shown in Fig.~\ref{fig:ssim-with-mome}. For cases without momentum corrections, their SSIM indices present relatively smaller values. Their error bars remain unchanged after removing bright spots. Once the momentum correction is operated, the mean of SSIM indices increases significantly, particularly using bright spot removal treatment. Accordingly, the momentum correction improves the quality of reconstructed images. Bright spot removal is required during momentum corrections.

\subsection{The equi-distance scheme for a muon momentum measurement system} \label{sec:result:MIS}

% \begin{figure}[htp]
%     \includegraphics[width=0.48\textwidth]{fig-muon-mome-seperation}
%     \caption{
%         The equiv-momentum-position(EMP) scheme of the multi-layer Cherenkov spectrometer. 
%         % The multi-layer Cherenkov detector's uniform position-resolution scheme is illustrated as follows: 
%         (a) the momentum distribution of cosmic muons.
%         (b) the scheme in Bae's original paper, with threshholds set at $[0.1, 1, 2, 3, 4, 5]\
%         \unit{GeV}$.
%         (c) EMP scheme of six detectors, with the lowest (\SI{0.1}{GeV}) and highest (\SI{5}{GeV}) thresholds
%         fixed as the same as Bae's scheme, and the intermediate layer's thresholds equally spcaded in the
%         magnitude of momentum.
%         Panel (d) same as (c) but with totally four detectors instead of six.
%         %
%     }
%     \label{fig:Bae-scheme-diag}
% \end{figure}

Measuring muon momentum presents considerable challenges due to their high average momenta ($\approx\,\,\SI{3}{\GeV/c}$), resulting in velocities extremely close to the speed of light. Traditional approaches for
measuring muon momentum typically involve either large-scale strong magnetic field
configurations that measure the muon momentum through trajectory deflections, or time-of-flight measurements that encounter
fundamental limitations due to the impractical timing precision required for particles moving at near-light
speed.
Recently, Bae et al. proposed an innovative approach that utilizes Cherenkov radiation to detect whether a
muon's velocity exceeds the light speed in a medium \cite{Bae2024}. This method employs a detector
system with multiple layers containing different materials or gas media at varying pressures, where each layer
has a distinct phase velocity of light. By passing through this layered structure, muons can be categorized
into specific momentum ranges based on which layers trigger Cherenkov radiation.
While this technique provides momentum range discrimination rather than precise momentum values, it offers
significant advantages by eliminating the need for magnetic fields and requiring minimal spatial footprint.
The method shows particular promise for cosmic muon scattering imaging applications, where relative momentum
determination serves as an effective auxiliary for imaging purposes. The practical feasibility of this
approach makes it especially valuable for field implementations where conventional methods would be
impractical.

\begin{figure}
	\centering
	\includegraphics[width=0.5\textwidth]{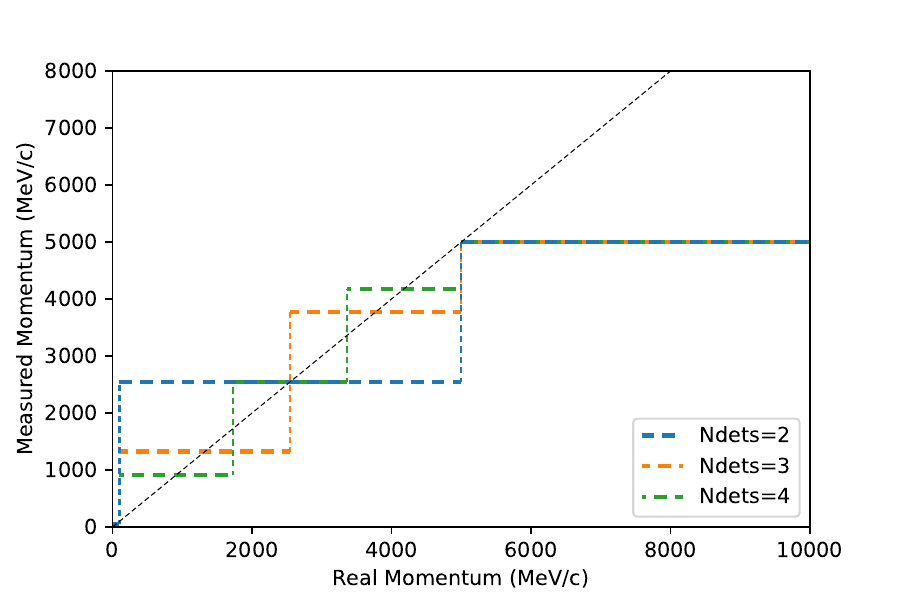}
	\caption{An example showing the equi-distance scheme of momentum measurement.
		The measured momentum versus real momentum under the equi-distance scheme using 2, 3, or 4 detectors.
		The lowest and highest thresholds are \SI{100}{MeV/c} and \SI{5}{GeV/c}, respectively.}
	\label{fig:equiv-magnitude-scheme}
\end{figure}

An equi-distance scheme for muon momentum discretion is proposed as illustrated in Fig.~\ref{fig:equiv-magnitude-scheme}. The real momentum in the region $(\SI{0.1}{GeV/c},~\SI{5}{GeV/c})$ is equally segmented. The measured momentum can be expressed by
\begin{align}
    p_i = \qty(0.1+ 4.9\frac{i-1}{N-1})~\unit{GeV/c}, 
\end{align}
where the integer, $i=1,2,\ldots,N$, is the detector layer order, and $N$ is the total number of detector layers. For every momentum segment in $(\SI{0.1}{GeV/c},~\SI{5}{GeV/c})$, the measured momentum equals the middle value of the real momentum. For muons with their momenta exceeding 5 GeV/c, the measured momentum is fixed to 5 GeV/c.

\begin{figure}
    \centering
    \includegraphics[width=0.5\textwidth]{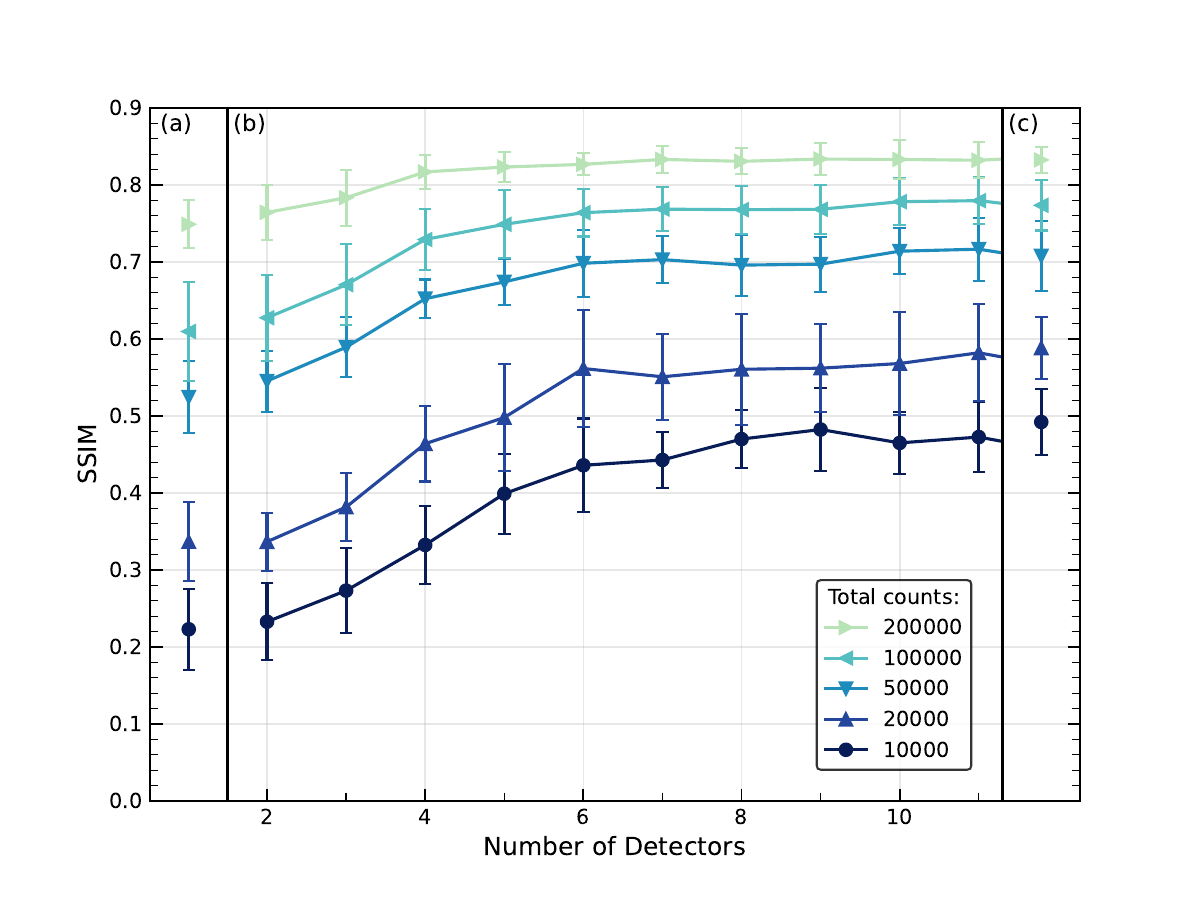}
    \caption{ (a) SSIM indices without momentum corrections. (b) Momentum corrected SSIM indices as a function of detector numbers. (c) SSIM indices corrected with real muon momentum. The error bars in all panels are estimated from different random seeds simulated separately.}
    \label{fig:sim-N-intervals}
\end{figure}

Figure \ref{fig:sim-N-intervals} presents the SSIM indices with respect to the number of detector
layers. As shown in Fig.~\ref{fig:sim-N-intervals}(b), the SSIM index increases quickly with less
than 6 detectors, and almost saturates afterwards. This varying trend is more pronounced with lower total counts. Adding more total counts into a simulation results in better imaging qualities.
Comparing Fig.~\ref{fig:sim-N-intervals}(a), (b) and (c), we can conclude that: 1) the imaging quality can be enhanced with even only 2 detector layers, 2) 6 detectors are sufficient to achieve the optimal SSIM in a given total count, 3) precise muon momentum measurement is not required in momentum corrected MST imaging.

\subsection{The equi-percentage scheme for a muon momentum measurement system}
\label{sec:result:MIS_custom}

\begin{figure}
	\includegraphics[width=0.5\textwidth]{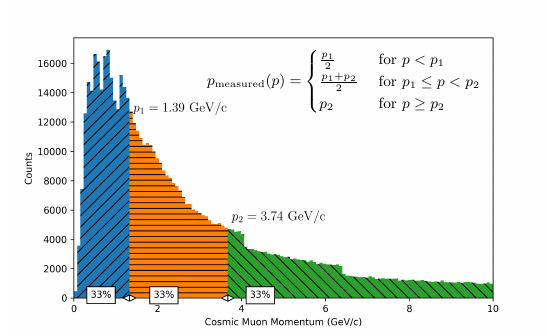}
	\centering
	\caption{An example showing the equi-percentage scheme of momentum measurement. The muon momentum spectrum is equally divided to three sections with two boundaries located at \SI{1.39}{\GeV/c} and \SI{3.74}{\GeV/c}, respectively. The step equation in the panel expresses the calculation of measured momentum, $p_{\rm measured}$.}
	\label{fig:muon-percentage}
\end{figure}

Different from Section~\ref{sec:result:MIS}, a equi-percentage scheme was proposed for muon momentum estimation in this subsection. As shown in Fig.~\ref{fig:muon-percentage}, this new scheme equally divides the probability distribution into several patches. When using two detectors, the triggering thresholds for each layer are set at \SI{1.39}{\GeV/c} and \SI{3.74}{\GeV/c}, dividing the momentum range into three intervals. Each interval
contains approximately one-third of the total muon population. Using the step function listed in Fig.~\ref{fig:muon-percentage}, the measured momenta for three sections are determined as \SI{0.695}{\GeV/c}, \SI{2.565}{\GeV/c} and \SI{3.74}{\GeV/c}, respectively. Similarly, when using $N$ detector layers, the thresholds are set to divide the muon momentum distribution into $N+1$ equal-percentage intervals. The measured momentum can be calculated as 
\begin{align}\label{equi-percentage}
   p_{\rm measured} = 
    \begin{cases}
        p_1/2, &  p < p_1 \\
        (p_1+p_2)/2, & p_1 \leq p < p_2  \\
        ... \\
        (p_{N-1}+p_N)/2, & p_{N-1} \leq p < p_N \\
        p_N, & p \geq p_N
\end{cases}
\end{align}
Herein, $p_N$ is less than 10 GeV/c. This scheme takes advantage of the fact that the cosmic muons are not equally distributed in
different ranges of momenta. Hence, placing too many detectors in regions with fewer muons would be wasteful, while placing too few detectors in momentum regions with a high density of muons would result in insufficient resolution for those momentum ranges. 

\begin{figure}
	\centering
	\includegraphics[width=0.5\textwidth]{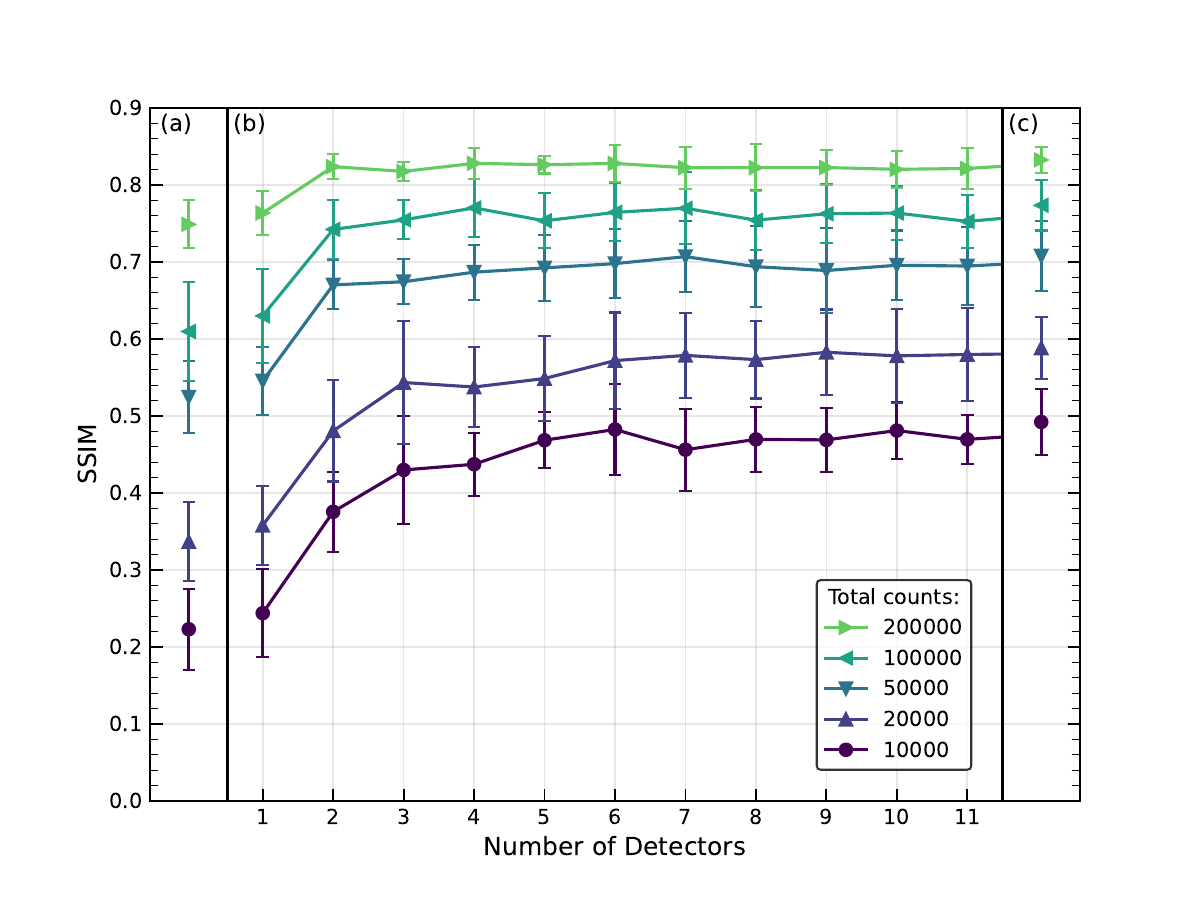}	
	\caption{(a) SSIM indices without momentum corrections. (b) Momentum corrected SSIM indices as a function of detector numbers. (c) SSIM indices corrected with real muon momentum.}
	\label{fig:NofInterval_vs_SSIM}
\end{figure}

Figure \ref{fig:NofInterval_vs_SSIM} presents the SSIM indices with respect to the number of detector layers. Using two detectors shows large enhancement in SSIM compared with no momentum corrections (the SSIM index increases from 0.22 to 0.38 for 10,000 events). With four detector layers, the SSIM index reaches 0.46 for 10,000 events, approaching the SSIM value of 0.49 achieved with fully accurate momentum and surpassing the SSIM of 0.44 obtained with six detector layers in the previous scheme. Compared to the equi-distance scheme, the equi-percentage scheme is highly advantageous in reducing the number of detector layers while maintaining good performance. Accordingly, only 2 $\sim$ 4 detector layers are required with sufficient imaging qualities.

\section{Discussion}
\label{sec:discussion}

\begin{figure}
 	\centering
 	\includegraphics[width=0.5\textwidth]{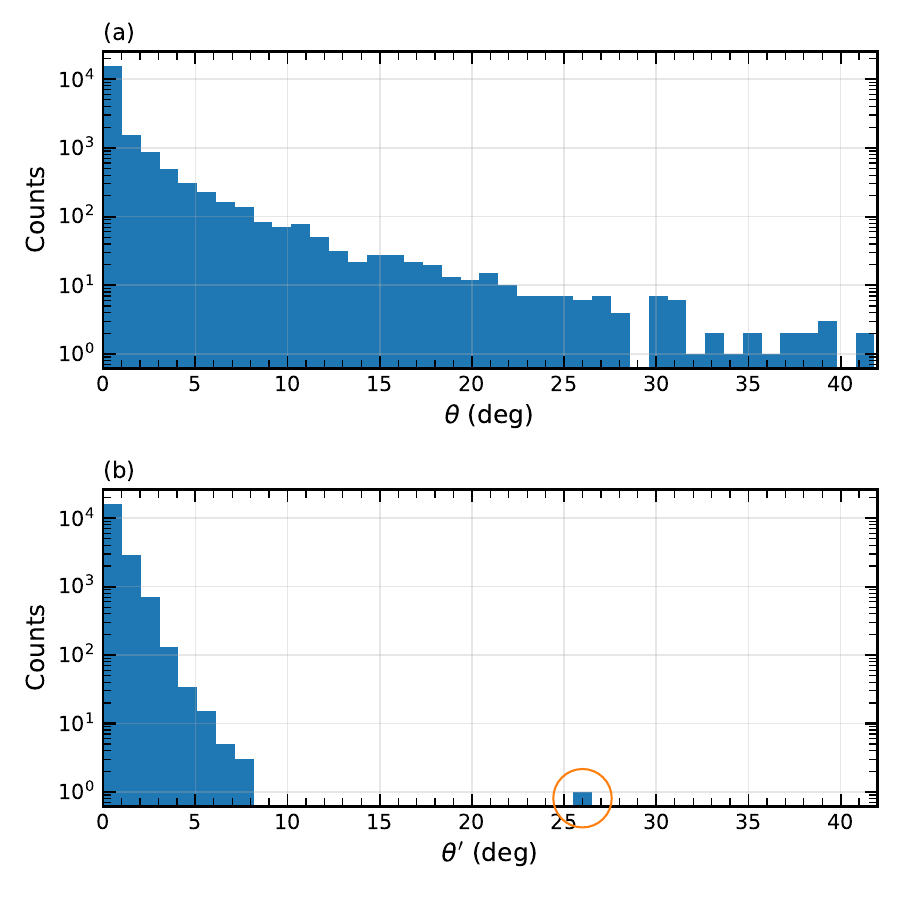}
 	\caption{
 		The angular distribution (a) before and (b) after momentum corrections. The circled event circled in panel (b) corresponds to the bright spot marked in Fig.~\ref{fig:result1}(f).
 	}
 	\label{fig:angle-distribution}
 \end{figure}

The results presented in this paper demonstrate that imaging quality improves significantly after
introducing momentum corrections, particularly in short-duration imaging, where the enhancement is especially pronounced. Additionally, we analyzed the statistical distribution of scattering angles before and after corrections as shown in Fig.~\ref{fig:angle-distribution}. Before corrections, the angular distribution spanned a wide range, extending beyond 40 degrees in Fig.~\ref{fig:angle-distribution}(a). This wide angular expansion became more concentrated after momentum corrections, evidenced by a steeper drop on the right side of the distribution curve in Fig.~\ref{fig:angle-distribution}(b). In companion with momentum corrections, individual events occur with very large corrected scattering angles as marked in a circle. Such large scattering angles contribute to a specific voxel during image reconstruction, causing its reconstructed value to become abnormally high --- far exceeding the values of other pixels in the image, shown as the circle brightest spot in Fig.~\ref{fig:result1}(f). According to Eqs.~(\ref{eq:tau_def}-\ref{eq:lambda_prime}), scattering angles are squared before being summation. This squaring operation further amplifies the contribution of events with large scattering angles, exacerbating the intensity of these anomalous pixels. 
 
The improvement in imaging quality using muon momentum corrections can be analyzed as follows. In MST, for monoenergetic muons, the primary component of their scattering angle distribution should follow a Gaussian distribution. However, the energy of cosmic muons distributes continuously in a wide range. Therefore, the scattering angle distribution of cosmic muons after passing through a material is no longer described by a simple Gaussian distribution function. The new angular distribution can still represent the properties of the imaged material. In principle, given sufficient measuring time and large enough samples, the variance of this distribution could be determined through averaging. However, the convergence rate of the sample variance depends on the concentration of the distribution
--- a more concentrated distribution converges faster, whereas a broader distribution requires more samples for the variance to stabilize. By measuring the muon momentum and applied it to correct the scattering angle, we effectively narrow the effective scattering angle distribution as shown in Fig.~\ref{fig:angle-distribution}(b). This also reduces the number of required muon events to achieve a stable variance estimation. Conversely, without momentum correction, the scattering angle distribution remains broad, leading to slower convergence as shown in Fig.~\ref{fig:angle-distribution}(a).  This explains why uncorrected images require significantly more muon events to suppress statistical fluctuations.
%and why they are more prone to outlier-induced artifacts (e.g., bright spots from rare large-angle scatters).
%
In summary, momentum-aware imaging accelerates statistical convergence by suppressing the
scattering angle distribution, enabling higher-quality imaging with fewer muons or shorter exposure times. This is particularly crucial for practical applications where data acquisition time is limited.

To measure the dispersion of the angular distribution, this paper considers the fourth-order moment --- also called kurtosis --- of the distribution. 
% At the same time, to eliminate the influence of scaling on the fourth-order moment, it is divided by the square of the variance (second order momentum).
The variable, kurtosis, is dimensionless that it is a fourth-order moment divided by the square of variance, to make it conserved under the change of unit or multiple of a constant.
The general formula for kurtosis, $\Kur$, is
\begin{align}
    \Kur(\theta) = \frac{\E (\theta - \E (\theta))^4}{[\E(\theta-\E(\theta))^2]^2},
    \label{eq:Kur-general}
\end{align}
where $\E$ denotes the expectation operation. Herein, the scattering angle distribution is centered at zero, and $\E(\theta) = 0$. The scattering angle in 3-D space $\theta \approx \sqrt{\theta_x^2 + \theta_y^2}$ is for the convenience of calculation.
% we multiply or divide the distribution by the same constant to normalize its variance (second-order moment). 
%Note that the mean of the scattering angle distribution is zero.
% Finally, the formula for this metric is:
The formula for muon scattering distribution is
\begin{align}
    \Kur \equiv  \frac{\frac{1}{N} \sum_{i=1}^{N} \theta_i^4} {\qty(\frac{1}{N}\sum_{i=1}^N \theta_i^2)^2} 
    = \frac{N\sum_{i=1}^{N} \theta_i^4} {\qty(\sum_{i=1}^N \theta_i^2)^2}. 
    \label{eq:Kur-of-theta}
\end{align}
Herein, $N$ is the total number of events, and $\theta_i$ represents the scattering angle of the $i$-th event in the three-dimensional space.
%Noting that $\theta \approx \sqrt{\theta_x^2+\theta_y^2}$ are directly used, since it is the same quality used in the imaging process in this paper.
$\Kur$ is a scale-independent  quantity that multiplying all pixel values by a constant does not affect its magnitude.  A smaller $\Kur$-value indicates a
more concentrated distribution, which is favorable for imaging.
% Additionally, consistent with the description in Section
% \ref{sec:methods:spots}, excluding the few events with the largest scattering angles ensures that
% the statistical range of this metric aligns with the actual imaging event range.

\begin{figure}
	\centering
	\includegraphics[width=0.5\textwidth]{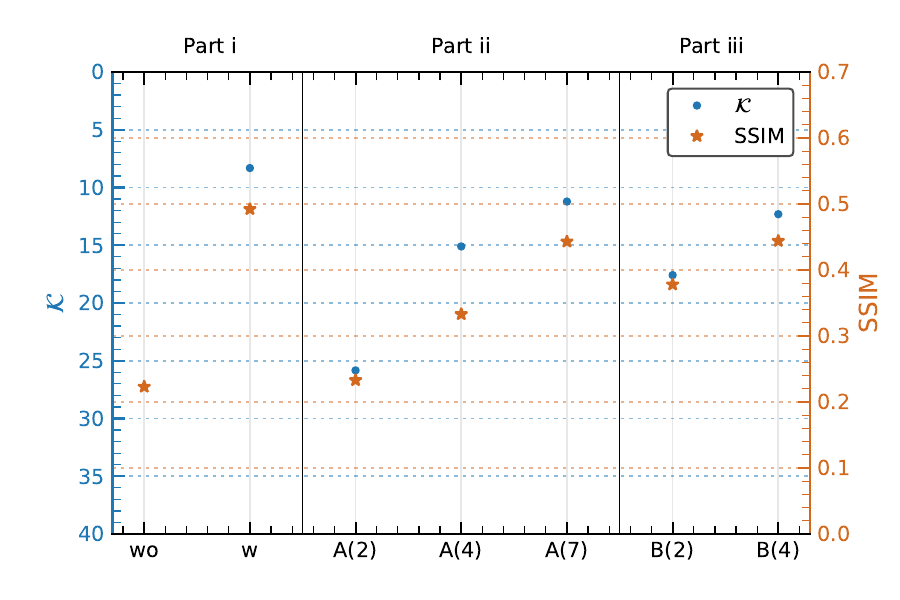}
	\caption{The kurtosis $\Kur$ of the scattering angle distribution (left axis) and the SSIM index of the reconstructed images (right axis). Three scenarios are presented in $x$-axis: (Part i) comparison between reconstructions with (precise muon momentum) and without momentum corrections; (Part ii) the equi-distance scheme (marked as A) with 2, 4 and 7 detectors; (Part iii)  the equi-percentage scheme (marked as B) with 2 and 4 detectors.}
	\label{fig:K-and-SSIM}
\end{figure}

Figure~\ref{fig:K-and-SSIM} presents the variation of $\Kur$ and SSIM index varying the momentum correction schemes. In Part i, the $\Kur$-value decreases from 36.2 to 9.45 after momentum corrections, demonstrating a significant improvement in the concentration of the
scattering angle distribution. In Parts ii and iii, both momentum measuring schemes exhibit progressively better image quality  with more detector layers, along with better concentration of scattering angles (i.e. lower $\Kur$). Thus, $\Kur$ shows a strong negative correlation with SSIM index. This relation confirms again that the imaging quality is improved with narrower angular distribution after using the momentum correction algorithm.

For a reference, imaging an object of single-material and single-thickness with mono-energy muons
gives a $\Kur$ of 2 calculated by Eq.~\eqref{eq:Kur-of-theta}
% \added{(see supplementary material for detailed derivation of this result)}
    % \deleted{\cite{Yu2025}}.
    \cite{Yu2025}.
% \pdfmargincomment{这里是附录，不是引文吧？}
This value presents the lower limit of $\Kur$ using any form of momenta corrections. The kurtosis introduced in this section can serve as an new indicator of imaging algorithms.

\section{Conclusion}
\label{sec:conclude}
This paper introduces a MST image reconstruction method utilizing momentum information. Under this
approach, the momentum-corrected imaging quality are significantly superior to those obtained
without momentum information. After incorporating momentum data, fluctuations in pixel values of
imaging noticeably decrease across most
regions. However, one or a few bright spots may appear in the image, greatly suppressing the
relative brightness of other areas and making the overall image quality appear suboptimal.
Nevertheless, this issue is effectively resolved by applying a post-processing method that removes
these few bright spots.
For momentum-range measurements based on Cherenkov radiation in muon imaging, this paper proposes an
improved interval-resolution positioning scheme. This method determines the medium for each layer of
the Cherenkov detector based on the percentile distribution of cosmic muon momentum, enabling more
efficient discrimination of natural muon momentum.
Under this scheme, just two layers of momentum detectors yield imaging quality significantly better
than that achieved without momentum information. With four detector layers, the results approach the
quality of imaging using precise momentum data, outperforming conventional schemes with uniformly
distributed momentum-resolution positions.
This study demonstrates that even a coarse momentum resolution using only two to four detector
layers can substantially enhance short-duration imaging quality. This underscores both the necessity
and feasibility of momentum detection in cosmic muon scattering imaging.
This method would also be promising for material identification in muon tomography, which will be
investigated in our future work.

\section{Acknowledgments}

This work is financially supported by the National Natural Science Foundation of China (Grant No. 12
105327, 12375292, 12405402, 12405337), the Guangdong Basic and Applied Basic Research Foundation (
Grant No. 2023B1515120067), and the State Key Laboratory of Nuclear Physics and Technology, Peking
University (Grant No. NPT2025KFY07).

This work is supported by  High Intensity heavy-ion  Accelerator Facility (HIAF) project approved by the National Development and Reform Commission of China.

This work is supported by the Research Program of 
State Key Laboratory of Heavy Ion Science and Technology, Institute of Modern Physics, Chinese
Academy of Sciences,
% Lanzhou 730000, China,
under Grant No. HIST2025CS06.

%\bibliography{momentum-mst.bib}
% \bibliography{mom-refs.bib}
% \onecolumn

\end{document}